\documentclass[twocolumn,showpacs,preprintnumbers,amsmath,amssymb]{revtex4}
\usepackage{graphicx}% Include figure files
\usepackage{subfigure} % Subfigures within figure
\usepackage{dcolumn}% Align table columns on decimal point
\usepackage{bm}% bold math
\begin{document}
%\draft

\title{Effective Time Reversal Symmetry Breaking and Energy Spectra of Graphene Armchair Rings}

\author{Tianhuan Luo}
\author{A.P. Iyengar}
\author{H.A. Fertig}
\affiliation{Department of Physics, Indiana University, Bloomington, IN 47405}
\author{L. Brey}
\affiliation{Instituto de Ciencia de Materiales de Madrid
(CSIC), Cantoblanco 28049, Spain}

\date{\today}

\begin{abstract}
We study the energy spectra and wavefunctions of graphene rings formed from
metallic armchair ribbons, near zero energy, to search for properties which
may be identified with ``effective broken time reversal symmetry'' (EBTRS).  Appropriately
chosen corner junctions are shown to impose phase shifts in the wavefunctions
that at low energies have the same effect as effective flux tubes passing near
the ribbon surface.  Closing the ribbon into a ring captures this flux and
yields properties that may be understood as signatures of EBTRS.  These include
a gap in the spectrum around zero energy, which can be removed by the application
of real magnetic flux through the ring.  Spectra of five and seven sided rings are also
examined, and it is shown these do not have particle-hole symmetry, which may
also be understood as a consequence of EBTRS, and is connected to the curvature
induced in the system when such rings are formed.  Effects of
deviations from the ideal geometries on the spectra are also examined.

\end{abstract}

 \maketitle

\section{Introduction}

Graphene is one of the most interesting  two dimensional
electron systems to become experimentally available in the last
decade \cite{Novoselov_2004}.  The low energy physics of this system
is dominated by two Dirac points per spin, through which the Fermi
energy passes when the system is nominally undoped \cite{Castro_Neto_RMP}.
Among the many fascinating ideas associated with
this system is a concept of ``effective time-reversal symmetry breaking'' \cite{Morpurgo_2006}.
There are two ways in which this occurs. Most directly, each individual Dirac point of
graphene does not reflect time-reversal symmetry by itself: time reversal
in this system involves interchanging the two valleys.  This can have dramatic
consequences.  For example, transport in graphene systems where disorder is ineffective
at intervalley scattering is expected to lead to weak {\it anti}-localization,
rather than localization \cite{Morpurgo_2006,Aleiner_2006,Morozov_2006,Ziegler_2006,Ostrovsky_2006,Tikhonenko_2008,Yan_2008}.

A second related effect occurs when a graphene surface has curvature.
The effective long-wavelength theory of the system in such cases include gauge fields which
can be understood as effective magnetic fields \cite{Vozmediano_1993,Lammert_2000,Morpurgo_2006}.
These fields are
directed in opposite directions for each valley, so that the system as a whole respects
time reversal invariance.  If there is no source of short-range scattering in the
system, such that intervalley scattering is negligible, the electrons can behave as
if they are moving in a non-uniform magnetic field \cite{Morpurgo_2006}.  Such a viewpoint
can be used to infer the single particle energy spectrum of a buckeyball \cite{Vozmediano_1993},
or in the vicinity of a disclination in an otherwise perfect graphene lattice \cite{Lammert_2000}.

Clearly, in the absence of applied magnetic fields and/or magnetic impurities, and if interactions
may be neglected, then the properties of graphene will be time-reversal invariant.  Nevertheless,
one may ask what properties can be understood as reflecting {\it effective} broken time-reversal
symmetry in an idealized situation.  One way to do this is by studying quantum rings.  In an
applied magnetic field, a generic one-dimensional system formed into a ring carries a
persistent current \cite{Imry_book}.  Systems with broken time reversal symmetry may
also carry spontaneous currents in zero magnetic field, as may be the case for example in
SrRu0$_4$ \cite{Kallin_2009}.  One can thus examine the low-energy electronic structure
of a graphene ring to see what properties it may have in common with systems where
time reversal symmetry is genuinely broken.  This is the subject of our study.

Graphene rings have been studied by several groups.  Nanoribbons closed into short nanotubes
have spectra which are sensitive to the precise boundary termination of the ring, and
may behave in a complicated way as the width is varied \cite{Nakamura_2004,Yoshioka_2008}.  Some studies
adopt simplified boundary conditions allowing for studies using the Dirac equation, so
that one may construct spectra from those of the individual valleys \cite{Recher_2007},
or treat many-body effects \cite{Abergel_2008}.  Tight-binding studies of flat graphene
rings involve different edge terminations and corner
geometries \cite{Recher_2007,Bahamon_2008,Wurm_2009a, Wurm_2009b},
and reveal spectra which are sensitive to both, consistent with previous studies
of transmission through polygons and junctions \cite{Iyengar_2008,Katsnelson_2008}.  
Some experimental
studies have recently been reported \cite{Russo_2008,Molito_2009}, in which Aharonov-Bohm oscillations
in graphene rings are reported.  This demonstrates that sufficiently
large phase coherence lengths may be reached
to allow quantum coherence effects to be observed, although current sample geometries
are not sufficiently controlled to allow direct comparison with the theoretical studies
of idealized models.

In this work, we will focus on graphene rings constructed from metallic armchair ribbons.  The
motivation for this choice is that it offers clear contrasts from what is expected
if the unique properties of its two valleys and sublattices are not properly accounted for.
Another important motivation for this choice is that 60$^\circ$ corner junctions may
be constructed for such ribbons which are perfectly transmitting \cite{Iyengar_2008} in
the lowest transverse subband.
Under such circumstances one may expect a conventional semiconductor to have a state at
zero energy, since the transverse confinement energy vanishes and backscattering
at the junctions is absent.  For graphene, we shall see that while there is no backscattering,
there is a phase shift upon passing through such junctions which open a gap around
zero energy even in the absence of disorder.  This phase shift is highly
analogous to one associated with the holonomy arising in a ``cut and paste''
procedure used to create a disclination; in particular it may be
modeled by introducing a (singular)
gauge potential \cite{Vozmediano_1993,Lammert_2000} which, in the present case,
induces an effective flux contained in ring.
Thus the repulsion of the spectrum away from zero energy in the absence of any real
magnetic flux
threading the ring may be understood as a manifestation of
``effective time-reversal symmetry breaking'' (ETRSB).

\begin{figure}[htb]
\subfigure[]{
\includegraphics[scale=.5]{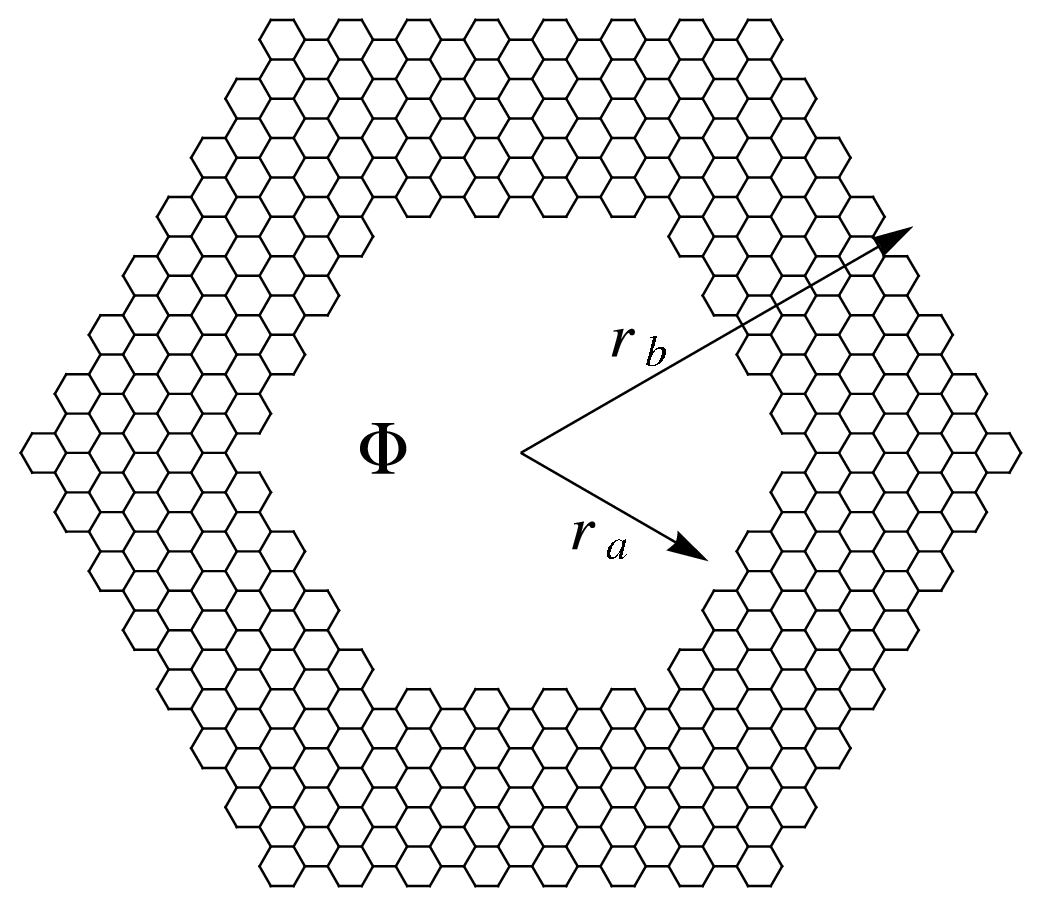}
\label{HexRingAC}
}
\subfigure[]{
\includegraphics[scale=.5]{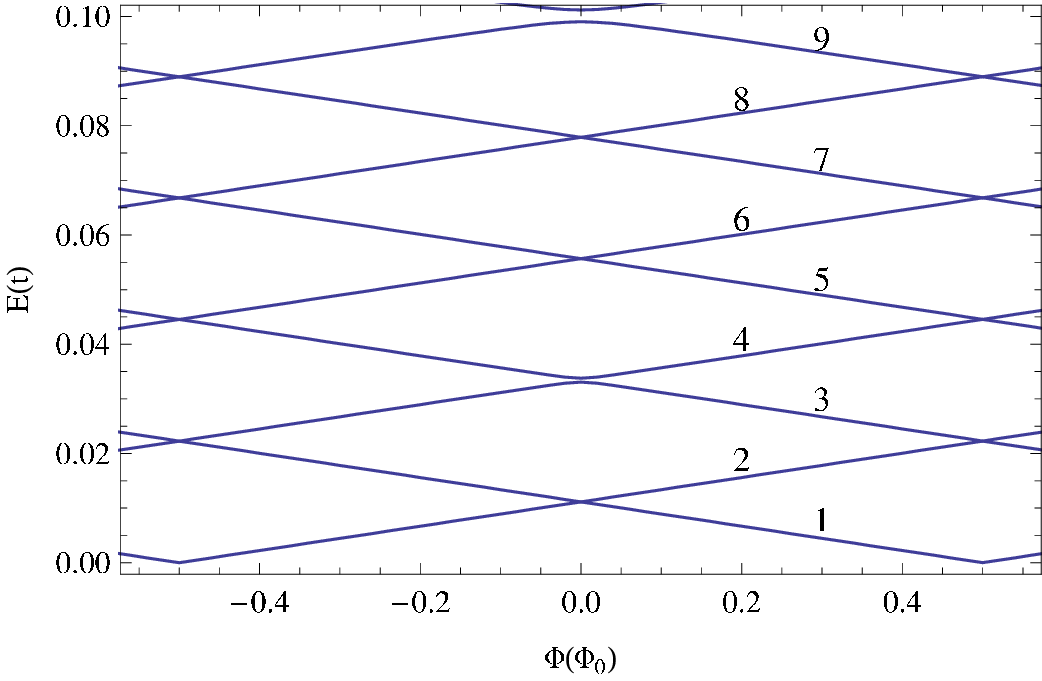}
\label{HexRingAC_spectrum}
}
\caption{Armchair hexagonal ring. (a) Illustration of a ring with inner radius $r_a$,
outer radius $r_b$, and flux $\Phi$ passing through the hole.  (b) Energy spectrum near
$E=0$ in units of the hopping matrix element t, for $E >0$.  For this ring the spectrum
is particle-hole symmetric, and  $r_a=32.5a$, $r_b=38.5a$.  Integers label the
energy levels.}
\end{figure}

Fig. \ref{HexRingAC} illustrates a simple geometry, a hexagonal ring constructed of metallic armchair nanoribbons,
and the associated energy spectrum as a function of magnetic flux through the hole of the
ring is presented in Fig. 1(b).  As we shall see, the spectrum for states closest to
zero energy may be understood in detail by accounting for the phase jumps at the junctions
as well as the six-fold symmetry of the system.  Levels with different rotational quantum numbers
cross (without level repulsion)  at higher energies in the absence of perturbations that
allow them to admix.  (Similar behavior has been observed previously for related
ring geometries \cite{Recher_2007,Bahamon_2008}.)  Studies \cite{Recher_2007} of zigzag hexagonal rings
have noted that their energy levels may be interpreted in terms of overlaid spectra for
each valley, each of which is asymmetric with respect to real magnetic flux through the ring, but
which together yield a spectrum which is symmetric with respect to time reversal.
Although states of armchair ribbons do not have valley index as a good quantum number,
we shall see that there is an analogous quantum number that can be associated with the
states of the lowest subband, so that in this case too the spectrum may be understood
as overlaid spectra which individually are not time-reversal symmetric.

Curvature may be introduced by considering rings with fewer or greater than six sides.
Such systems may be related to hexagonal rings by a cut and past procedure in which
one either cuts out one or more sides of the ring, and stitches together the dangling
bonds, or cuts open the system at a corner and inserts one or more extra sides.
This is highly analogous to the procedure by which one creates a disclination in
a perfect graphene sheet \cite{Vozmediano_1993,Lammert_2000}.
Since the procedure results in changing the number of corners, the effective flux
through the ring changes, modifying the
spectrum near zero energy in a way which may be understood in detail.
A prominent feature of the spectra of such rings is that they are not particle-hole symmetric.
In general, one may show that a honeycomb lattice in which closed loops always
have an even number of steps must have a particle-hole symmetric spectrum.
Interestingly,
for five-fold and seven-fold rings, any small loop must have an even
number of steps, but loops that surround the hole of the ring will be
odd in number.  Thus the breaking of particle-hole symmetry is associated
with a {\it topological} property of the ring, and is inextricably tied to
the curvature induced in forming pentagons and heptagons from graphene.  Thus
the broken particle-hole symmetry in the spectrum may be understood as
another manifestation of EBTRS.

Clearly, to observe these effects directly requires extreme
control of the lattice,
beyond the capabilities of current fabrication technologies.
We briefly examine the effects of disorder and perturbations
near the corners and edges to see how disturbing these are to
the features described above.  We find that the most profound
changes in the spectrum occur when the corners are modified from
their ideal form: in particular such defects can induce zero
energy states which are absent in a perfect ring geometry.
These are localized states which are insensitive to the
flux through the ring, and so do not induce a persistent
current.  Perturbations near the edges shift the spectrum
and may contribute nearly localized states, but do not in
general bring states to zero energy at zero flux,
so that the main feature which may be ascribed to EBTRS remains.

This article is organized as follows.  In Section \ref{section:prelim} we
discuss the subband structure of the armchair ribbon, and the phase jump
associated with creating a 60$^\circ$ corner in such a ribbon.
Section \ref{section:numerics} describes our results
for ideal rings, and we demonstrate that they can be understood
in great detail from the results of the previous section.
In Section \ref{section:perts} we examine the effects of a few
deviations from the perfect ring geometry, and finally
we conclude with a summary in Section \ref{section:summary}.

\section{Metallic Armchair Nanoribbons and the Transmitting Corner Junction}
\label{section:prelim}

\begin{figure}[htb]
\includegraphics[scale=.5]{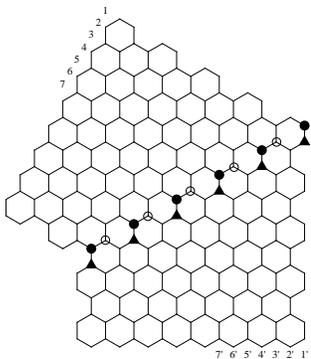}
\caption{Geometry for an armchair nanoribbon corner junction.  Note that
atoms from both sublattices appear
with equal frequency at the edge of each ribbon, implying vanishing boundary conditions for both.
Transverse modes of the ribbons may be matched at the junction by equating wavefunctions
on open and closed circles, and on triangles.
The integers spanning the width of the incoming
and outgoing ribbons correspond to the index $\ell$ in Eq. \ref{overlap}.}
\label{AC_ribbon}
\end{figure}

Figure \ref{AC_ribbon} illustrates two armchair nanoribbons meeting
at 60$^{\circ}$ corner junction.  A careful
comparison of the wavefunctions and energy spectra of armchair nanoribbons as found from
numerical tight-binding calculations and from analytical solutions of the
continuum model based on the
Dirac equation demonstrates that these may be brought into quantitative agreement,
provided one adopts vanishing boundary conditions for both the A and B sublattices
at the edges of the ribbon \cite{Brey_2006b}.  In the continuum this means
that the boundary condition admixes states from the two valleys around the Dirac points \textbf{K}
and \textbf{K}$^{\prime}$.  This is in contrast to the situation for zigzag
edges, for which the wavefunctions vanish on only one of the two sublattices,
and there is no valley mixing due to boundary conditions \cite{Brey_2006b}.
Moreover, in the lowest transverse subband for a zigzag nanoribbon, the direction
of current is associated with a particular valley.
This chirality underlies proposals to use zigzag graphene nanoribbons
as elements in ``valleytronics'' applications \cite{Rycerz_2007}.
Because armchair ribbons admix valleys one may suppose that such physics
is irrelevant to them.  However, armchair nanoribbons turn out to support an analogous chirality,
as we now demonstrate.

In the continuum limit, the positive energy wavefunctions
of an armchair nanoribbon may be written in the form
\begin{widetext}
\begin{equation}
\psi_{p_x,p_y}(x,y)={1 \over {\sqrt{2W}}}\left\{ \left(
\begin{array}{c} 1 \\
\frac{p_x+ip_y}{p}
\end{array}
\right)e^{i{\bf K}\cdot{\bf r}}e^{ip_xx}
-
\left(
\begin{array}{c} 1 \\
\frac{p_x+ip_y}{p}
\end{array}
\right)e^{i{\bf K}^{\prime}\cdot{\bf r}}e^{-ip_xx}
\right\}
e^{ip_yy},
\label{arm_ribbon}
\end{equation}
\end{widetext}
where the upper (lower) entry represents the amplitude on the $A$ ($B$)
sublattice, $W$ is the ribbon width, ${\bf K}=(-{{4\pi} \over {3a}},0)$,
${\bf K}^{\prime}=({{4\pi} \over {3a}},0)$, and $a$ is the lattice
constant.  The value of $p_x$ must be chosen such that the total
amplitude at the edges of the ribbons vanishes, so that $p_x \rightarrow p_n$
comes in quantized values \cite{Brey_2006b}.  For metallic ribbons, the lowest subband
satisfies $p_{n=0}=0$.  These wavefunctions have energy
$\varepsilon=v_F|{\bf p}|$, where $v_F$ is the speed of electrons
near the Dirac points.  Negative energy wavefunctions are related to those of
Eq. \ref{arm_ribbon} by changing $1 \rightarrow -1$ in the $A$ sublattice amplitudes.

The metallic case ($p_n=0$) allows an interesting simplification of the transverse
wavefunction, yielding
\begin{widetext}
\begin{equation}
\psi_{0,p_y}(x,y)={1 \over {\sqrt{2W}}}\left\{ \left(
\begin{array}{c} 1 \\ isgn(p_y)
\end{array}
\right)_{\bf K}
-
\left(
\begin{array}{c} 1 \\
isgn(p_y)
\end{array}
\right)_{{\bf K}^{\prime}}
\right\}
e^{ip_yy}.
\label{arm_ribbon_zero}
\end{equation}
\end{widetext}
The subscripts \textbf{K},\textbf{K}$^{\prime}$ are introduced in the above
expression to denote the fact that the spinors provide amplitudes for each of
the two valleys.  Writing the amplitudes for the wavefunction in vector form,
${\bf \Psi}_{p_n,p_y}=(\Phi^{A,{\bf K}*},\Phi^{B,{\bf K}*},\Phi^{A,{\bf K}^{\prime}*},\Phi^{B,{\bf K}^{\prime}*})^{\dag}$, the $p_n=0$ wavefunction is an eigenstate of the matrix
\begin{equation}
T=
\left(
\begin{array}{cccc}
0 & 0 & 0 & i \\
0 & 0 & -i & 0 \\
0 & i & 0 & 0 \\
-i & 0 & 0 & 0 \\
\end{array}
\right).
\end{equation}

The matrix $T$ interchanges the amplitudes between both sublattice and valley,
and multiplies by a phase $\pm i$ \cite{com_T}.  This particular matrix may be shown to
commute with the Hamiltonian in the continuum limit, so that it represents a
symmetry of graphene at low energies.
Moreover, the boundary condition for armchair edges does not violate this
symmetry.  Thus, any eigenstate of a graphene armchair nanoribbon may be
expressed as an eigenstate of $T$.  In general this requires admixing
states of different signs of $p_n$.  However for the special case of $p_n=0$,
one finds
\begin{equation}
T {\bf \Psi}_{0,p_y}=sgn(\varepsilon)sgn(p_y) {\bf \Psi}_{0,p_y}.
\label{T_eigenvalues}
\end{equation}
Thus, for metallic nanoribbons the eigenvalue of $T$ in the lowest subband ($p_n=0$)
is tied to the direction of current $sgn(p_y)$, in a way that is highly analogous
to the connection between current direction and valley index for zigzag
nanoribbons, with the eigenvalue of $T$ playing the role of valley index.

For our present study of armchair rings, the corner junctions as illustrated in
Fig. \ref{AC_ribbon} are used.  Such junctions were shown previously
to support perfect transmission \cite{Iyengar_2008}. The transmission
amplitude may be computed by matching wavefunctions and currents for the two
ribbons on the joining surface.  Current matching is in general inconvenient
because it involves products of wavefunctions.  This requirement
may be simplified for the lowest
subband, because the
wavefunctions
vanish on the open circles in Fig. \ref{AC_ribbon}.
Thus one only need match the currents on the bonds connecting the closed
circles to the triangles.  This may be accomplished straightforwardly
by matching the wavefunctions on the triangles as well.  For low energies,
in which we focus on the lowest transverse subband and
take the $p_y \rightarrow 0$
limit, referring to the labeling in Fig. \ref{AC_ribbon},
the resulting matching conditions are \cite{Iyengar_2008}
\begin{widetext}
\begin{equation}
\begin{array}{ccc}
\psi^{(A)}_{0,0}(x=1,y)=\psi^{(B)\prime}_{0,0}(x^{\prime}=1^{\prime},y^{\prime}) &
          \psi^{(B)}_{0,0}(x=2,y)=\psi^{(A)\prime}_{0,0}(x^{\prime}=1^{\prime},y^{\prime}) &
          \psi^{(B)}_{0,0}(x=3,y)=\psi^{(A)\prime}_{0,0}(x^{\prime}=3^{\prime},y^{\prime})\\
\psi^{(A)}_{0,0}(x=4,y)=\psi^{(B)\prime}_{0,0}(x^{\prime}=4^{\prime},y^{\prime}) &
          \psi^{(B)}_{0,0}(x=5,y)=\psi^{(A)\prime}_{0,0}(x^{\prime}=4^{\prime},y^{\prime}) &
          \psi^{(B)}_{0,0}(x=6,y)=\psi^{(A)\prime}_{0,0}(x^{\prime}=6^{\prime},y^{\prime})\\
\psi^{(A)}_{0,0}(x=7,y)=\psi^{(B)\prime}_{0,0}(x^{\prime}=7^{\prime},y^{\prime}) &
          \psi^{(B)}_{0,0}(x=8,y)=\psi^{(A)\prime}_{0,0}(x^{\prime}=7^{\prime},y^{\prime}) &
          \psi^{(B)}_{0,0}(x=9,y)=\psi^{(A)\prime}_{0,0}(x^{\prime}=9^{\prime},y^{\prime})\\
\cdot & \quad\quad \cdot & \quad\quad \cdot\\
\cdot & \quad\quad \cdot & \quad\quad \cdot\\
\cdot & \quad\quad \cdot & \quad\quad \cdot\\
\end{array}
\label{match-armchair}
\end{equation}
\end{widetext}
The transmission amplitude between states of
the ribbons can be computed in the single mode approximation
(SMA) \cite{Londergan_1999,Iyengar_2008} by computing the overlaps $M_{0,0}(p_y)$ of the
wavefunctions on the joining surface,
\begin{equation}
M_{0,0}(p_y) \equiv \int d\lambda \,\, \psi^{(1)}_{0,p_y}(x(\lambda),y(\lambda)) \,
\psi^{(2)*}_{0,p_y}(x(\lambda),y(\lambda)),
\label{M00}
\end{equation}
where $\psi^{(1)}_{0,p_y}$ is the wavefunction to the left of the junction,
$\psi^{(2)}_{0,p_y}$ is the wavefunction to the right of the junction,
and  $\lambda$ parameterizes the joining surface.
Note that in the limit $p_y \rightarrow 0$, there is no actual $y$  dependence in
$\psi^{(\mu)}_{0,0}(x,y)$.
From Fig. \ref{AC_ribbon} one may see that the positions
denoted as $x(^\prime) = \ell(^\prime)$ demarcate increments of length $a/2$.
The meaning of the formal expression (Eq. \ref{M00}),
using Eq. \ref{arm_ribbon},
then takes the form for $p_y>0$, $\varepsilon>0$,

\begin{widetext}
\begin{eqnarray}
&~&M_{0,0}(p_y \rightarrow 0) = \nonumber\\
\nonumber\\
&\mathop{\sum_{\ell}}& \Biggl\{
i\left[ \exp\left(-i\frac{4\pi}{3}\left({3 \over 2} \ell+{1 \over 2}\right)\right)
- \exp\left(i\frac{4\pi}{3}\left({3 \over 2} \ell+{1 \over 2}\right)\right) \right]
\left[ \exp\left(i\frac{4\pi}{3}\left({3 \over 2} \ell+{1 \over 2}\right)\right)
- \exp\left(-i\frac{4\pi}{3}\left({3 \over 2} \ell+{1 \over 2}\right)\right) \right]
\nonumber \\
&-&i\left[ \exp\left(-i\frac{4\pi}{3}\left({3 \over 2} \ell+1\right)\right)
- \exp\left(i\frac{4\pi}{3}\left({3 \over 2} \ell+1\right)\right) \right]
\left[ \exp\left(i\frac{4\pi}{3}\left({3 \over 2} \ell+{1 \over 2}\right)\right)
- \exp\left(-i\frac{4\pi}{3}\left({3 \over 2} \ell+{1 \over 2}\right)\right) \right]
\nonumber \\
&-&i\left[ \exp\left(-i\frac{4\pi}{3}\left({3 \over 2} \ell+{3 \over 2}\right)\right)
- \exp\left(i\frac{4\pi}{3}\left({3 \over 2} \ell+{3 \over 2}\right)\right) \right]
\left[ \exp\left(i\frac{4\pi}{3}\left({3 \over 2} \ell+{3 \over 2}\right)\right)
- \exp\left(-i\frac{4\pi}{3}\left({3 \over 2} \ell+{3 \over 2}\right)\right) \right]
\Biggr\}
\nonumber \\
&&\quad\quad\quad\quad\Biggr/\sum_\ell \Biggl\{
\left| \exp\left[i\frac{4\pi}{3}\left({3 \over 2} \ell+{1 \over 2}\right)\right]
- \exp\left[-i\frac{4\pi}{3}\left({3 \over 2} \ell+{1 \over 2}\right)\right] \right|^2 \nonumber\\
&+&\left| \exp\left[i\frac{4\pi}{3}\left({3 \over 2} \ell+{1}\right)\right]
- \exp\left[i\frac{4\pi}{3}\left({3 \over 2} \ell+{1}\right)\right] \right|^2
+ \left |\exp\left[i\frac{4\pi}{3}\left({3 \over 2} \ell+{3 \over 2}\right)\right]
- \exp\left[-i\frac{4\pi}{3}\left({3 \over 2} \ell+{3 \over 2}\right)\right] \right|^2 \Biggr\}
.\label{overlap}\\
\nonumber\\
\nonumber
\end{eqnarray}
\end{widetext}
The denominator in Eq. \ref{overlap} comes from normalizing the wavefunctions.
Eq. \ref{overlap} yields $M_{0,0}(p_y \rightarrow 0)=i$; more generally,
one may verify that $M_{0,0}(p_y \rightarrow 0)=i$ for $p_y\varepsilon>0$, and
$M_{0,0}(p_y \rightarrow 0)=-i$ for $p_y\varepsilon<0$.  The result confirms the
unit transmission found previously for these junctions \cite{Iyengar_2008}, and
also implies a remarkable property: the wavefunctions should jump by a factor
of $\pm i$ as one passes through a junction, with the sign determined by the
eigenvalue of the $T$ operator defined above.

This phase shift can also be viewed from the continuum perspective.
Consider a long metallic armchair nanoribbon segment with its front
and back ends identified to form a cylinder, effectively a short, fat nanotube.
We may simulate the phase shift associated with the junction shown in
Fig. \ref{AC_ribbon} by adding a gauge field. 
Our Hamiltonian has the form ($\hbar=1$)
\begin{equation}
H({\bf p})=v_F
\left(
\begin{array}{cccc}
0 & -p_x+ip_y & 0 & 0 \\
-p_x-ip_y & 0 & 0 & 0 \\
0 & 0 & 0 & p_x+ip_y \\
0 & 0 & p_x-ip_y & 0 \\
\end{array}
\right),
\label{H_0}
\end{equation}
where $p_{x(y)}={1 \over i}\partial_{x(y)}$.
Writing
${\bf \Psi}_{p_n,p_y}=e^{i\chi}{\bf \Psi}_{p_n,p_y}^{\prime}$, we can rewrite
the Dirac equation for the wavefunctions in the form
$H({\bf p}-{\bf A}){\bf \Psi}_{p_n,p_y}^{\prime} = \varepsilon {\bf \Psi}_{p_n,p_y}^{\prime}$,
with ${\bf A}=\partial_y\chi\,\hat{y}$ and $\chi={{\pi} \over 2}T\Theta(y)$, where
the junction between ribbons is located at $y=0$.  In this representation,
${\bf \Psi}_{p_n,p_y}^{\prime}$ is continuous across the junction, and the phase
jump is fully implemented by the $e^{i\chi}$ factor.

\begin{figure}[htb]
\center
\includegraphics[scale=.35]{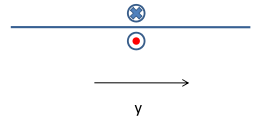}
\caption{({\it Color online.}) Geometry for flux tubes, arranged to produce the effects of the
phase factor created by the corner junction.  Effective flux above and below the plane of the
ribbon run in opposite directions.}
\label{flux_simple}
\end{figure}

The presence of the gauge field can be interpreted as being due to a pair of
solenoids carrying magnetic flux in opposite directions, one above the plane of the
ribbon, the other below.  In this way, one can understand the problem of $n$-sided
rings constructed from metallic armchair ribbons and corner junctions such as shown in
Fig. \ref{AC_ribbon} as being the same (near zero energy) as the problem of a ribbon
closed into a cylinder (i.e., a short nanotube), with $2n$ flux tubes, half threaded through 
in one direction and
half just outside it in the other direction.  In this way the system has properties
illustrating EBTRS.  Just as in examples of this effect in graphene
\cite{Vozmediano_1993,Lammert_2000},
{\it real} time reversal symmetry
is preserved for the system, in this case because the effective flux
runs in opposite directions for different eigenvalues of the operator $T$.
As we shall see in detail below, the low energy spectra of such
rings as found from solutions of the tight-binding model behave
precisely as if these phase jumps are present.  The resulting spectra present
properties which may be understood as signaling the EBTRS in graphene.

\section{Numerical Results}
\label{section:numerics}

The simplest ring system one can study using metallic armchair ribbons and the 60$^{\circ}$
corner junctions discussed above is the hexagonal ring, as illustrated in Fig. \ref{HexRingAC}.
The fact that the junctions are perfectly transmitting in the lowest
subband might lead one to think that the low energy spectrum is the same
as that of a metallic armchair ribbon closed into an cylinder (i.e.,
a very short carbon nanotube.)  If this were the case, one would expect
states at zero energy when no external magnetic flux threads
the ring.  Our discussion above however indicates that one needs to include
the effective flux passing through the ring to understand the spectrum.

Fig. \ref{HexRingAC_spectrum} is the spectrum obtained from computing the eigenvalues
of the tight-binding model near zero energy, as a function of flux $\Phi$ through
the ring.  Note in these calculations we include only the phase factors
in the hopping matrix elements due a solenoid passing through the hole
of ring; magnetic flux through the individual plaquettes of the honeycomb
lattice is not included.  In general
the spectrum in this case is perfectly
particle-hole symmetric, so only positive energy states are displayed.

The form of the spectrum is highly reminiscent of what has been seen previously
in hexagonal rings with zigzag edges \cite{Recher_2007}.  In particular it
takes the form of two sets of spectra, each with broken time reversal symmetry such that
the energies have a particular sign of slope near $\Phi=0$.  The spectra are
effectively time-reverses of one another so that the spectrum as a whole
has time-reversal symmetry; in particular the spectrum evolves in the same
way whether positive or negative flux is threaded through the hole.  In the
case of zigzag ribbons the two sets of spectra are associated with the two
valleys.  In the present case, they are associated with the two eigenvalues
of the matrix $T$.  Note that the crossing of the energy states through zero
at $\Phi=\pm\Phi_0/2$ may be understood as resulting from the sum of the
effective fluxes due to the corner junctions and that of the real field
summing to an integral number of flux quanta.

\begin{figure}[htb]
\includegraphics[scale=.75]{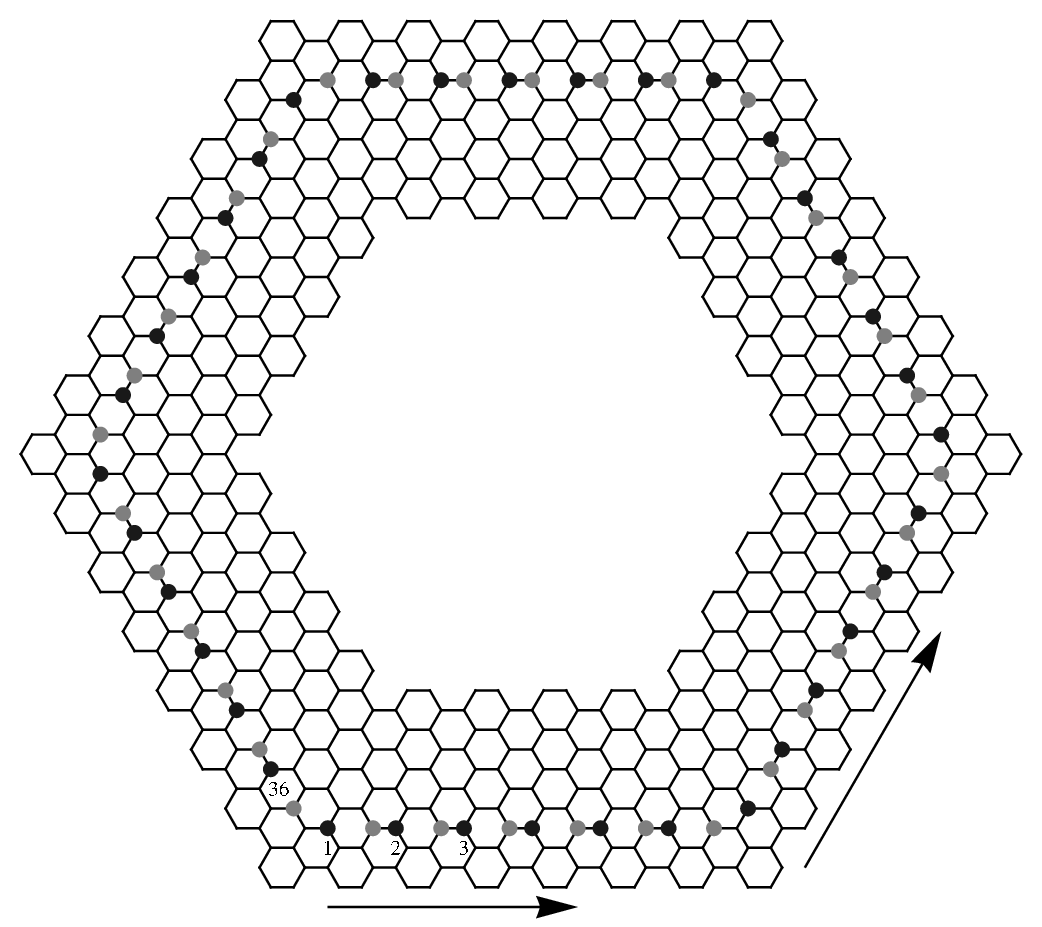}
\caption{Hexagonal ring illustrating site labels for examination of wavefunction.
Black dots indicate sites on $A$ sublattice, gray dots are on $B$ sublattice.
The actual ring used in calculations has $r_a=32.5a$ and $r_b=38.5a$, with $a$ the lattice
constant of the underlying triangular lattice.}
\label{HexRing_labeled}
\end{figure}

\begin{figure}[htb]
\subfigure[n=1; A sublattice]{
\includegraphics[scale=.35]{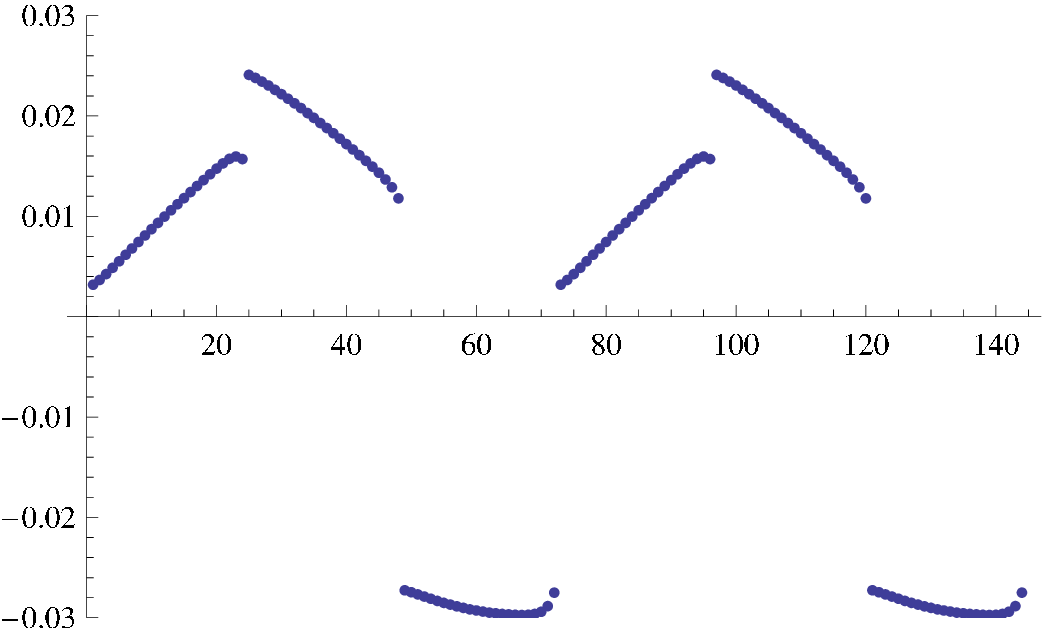}
}
\subfigure[n=2; A sublattice]{
\includegraphics[scale=.35]{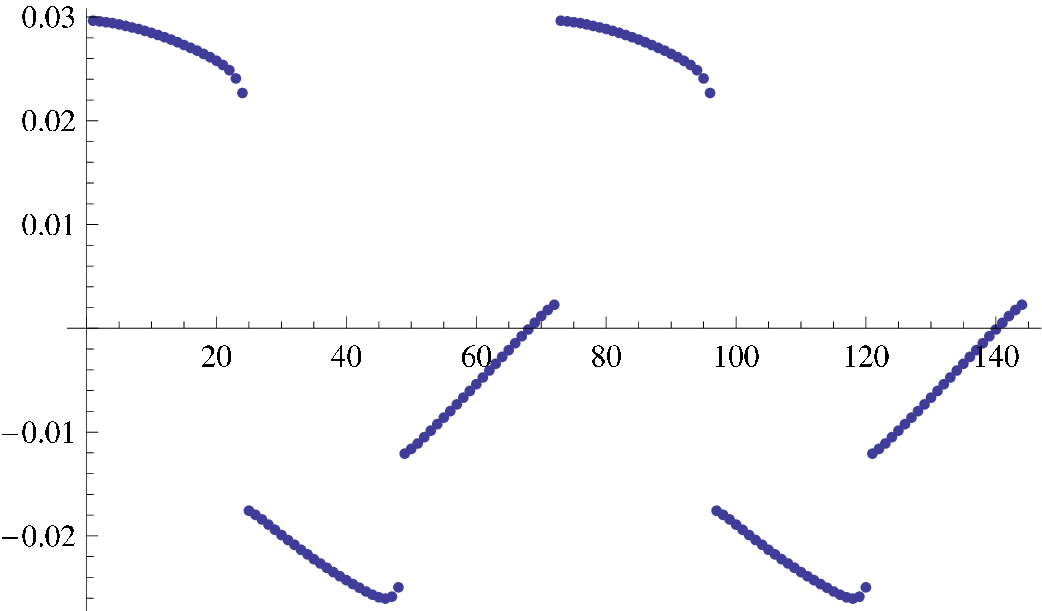}
}
\subfigure[n=3; A sublattice]{
\includegraphics[scale=.35]{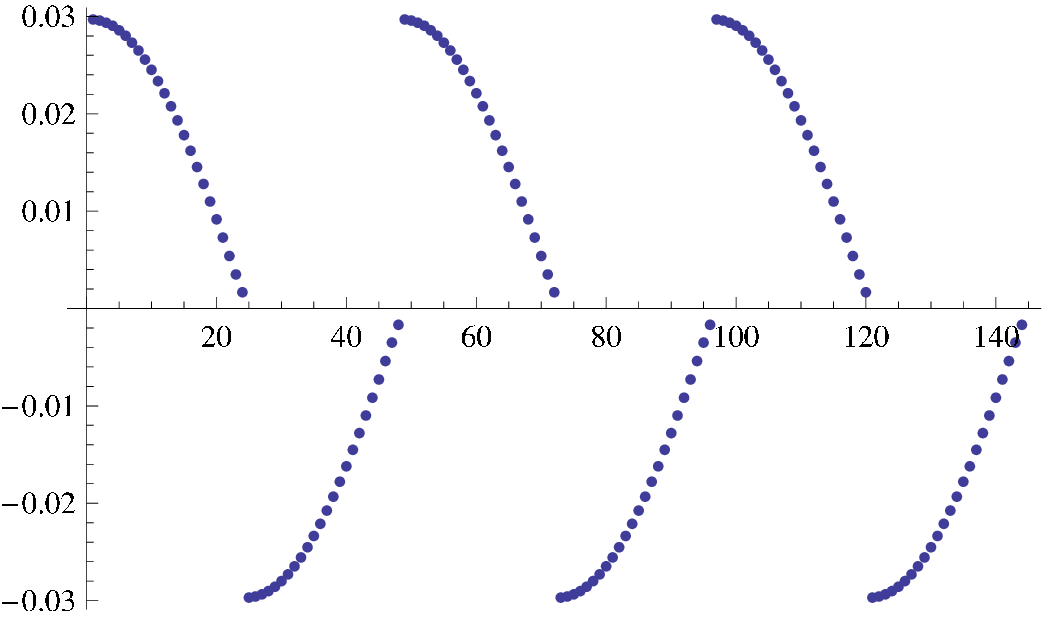}
}
\subfigure[n=4; A sublattice]{
\includegraphics[scale=.35]{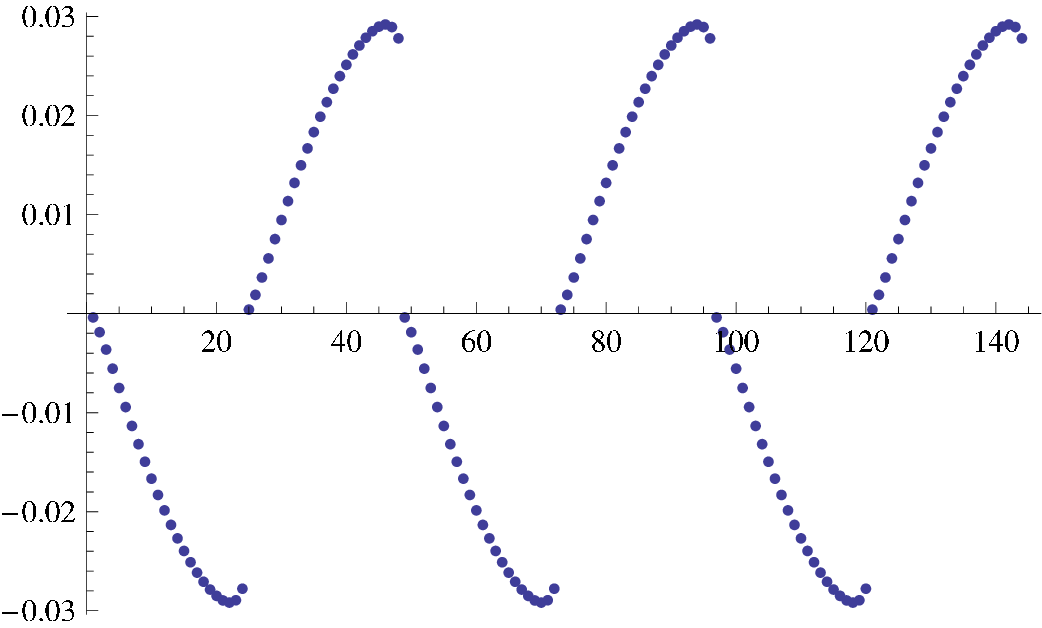}
}
\caption{({\it Color online.}) Wavefunctions  at zero flux $\Phi$ for the four lowest eigenstates $n=1,2,3,4$
illustrated in Fig. \ref{HexRingAC_spectrum}.}
\label{HexRing_spectra}
\end{figure}

\begin{figure}[htb]
\subfigure[n=1; $A$ sublattice]{
\includegraphics[scale=.35]{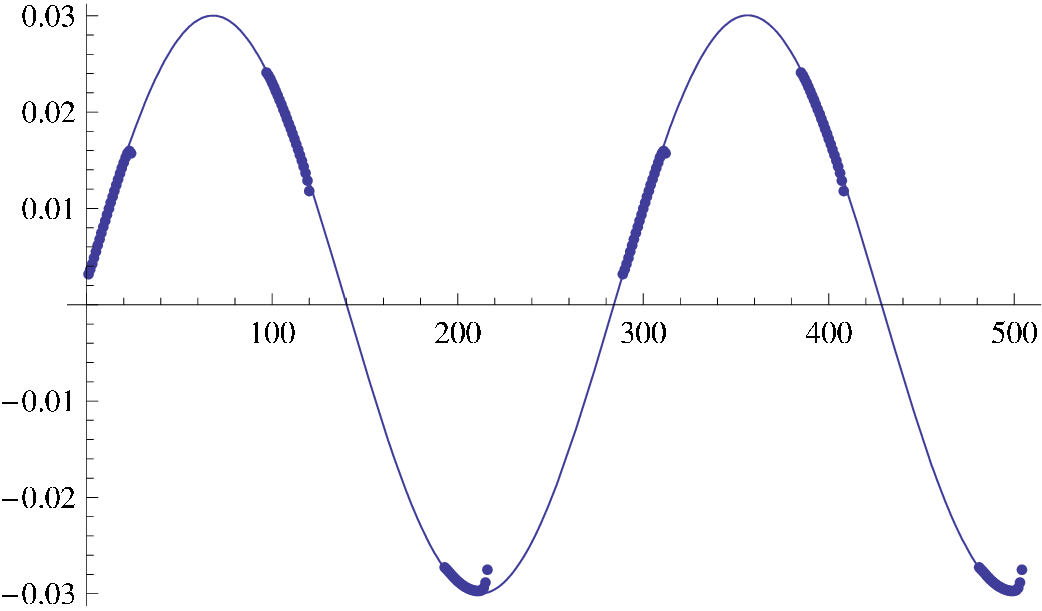}
}
\subfigure[n=3; $A$ sublattice]{
\includegraphics[scale=.35]{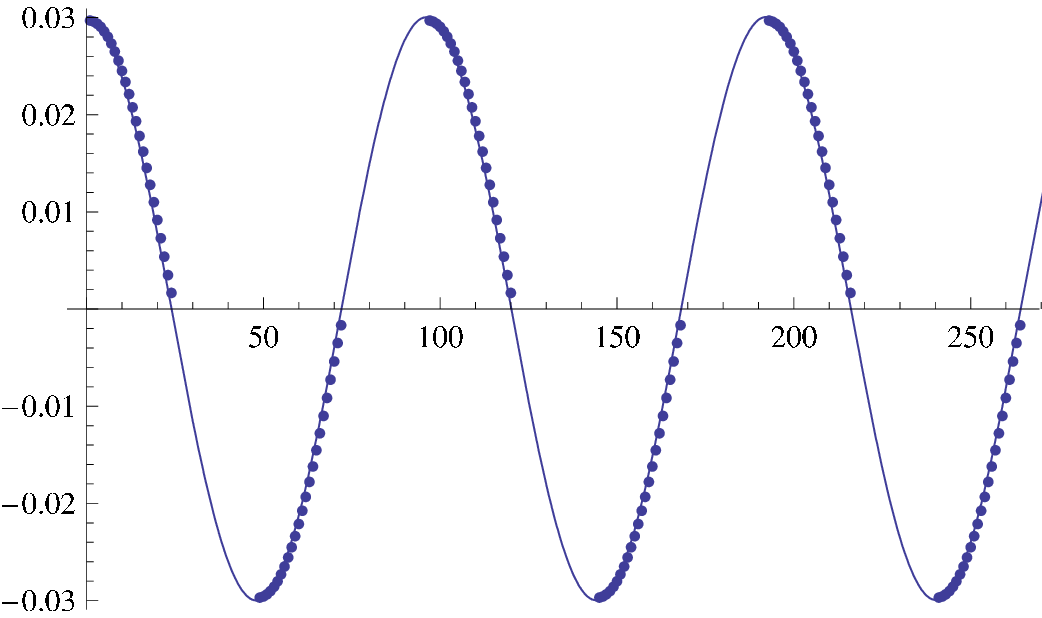}
}
\caption{({\it Color online.}) Wavefunctions at zero flux $\Phi$ for the $n=1,3$ eigenstates with phase jumps added (see text) and fitted to sine functions.
}
\label{HexRing_spectra_fit}
\end{figure}

The phase jumps associated with the corners may be demonstrated explicitly by
a careful examination of wavefunctions.  Fig. \ref{HexRing_labeled} labels a set
of sites around the ring, and Fig. \ref{HexRing_spectra} illustrates the
wavefunctions
on one of the sublattices for the four lowest positive energy levels at $\Phi=0$.
The jumps in amplitude associated with passing through the corners of the junctions
are quite apparent.  Moreover, the amplitudes in the sides of the ring
reflect the $e^{ip_yy}$ phase
factor in Eq. \ref{arm_ribbon} very accurately,
provided one recognizes that diagonalization of the tight-binding Hamiltonian results
in real eigenvectors, so that our wavefunctions are actually linear combinations
of states with positive and negative values of $p_y$.
To see this, we fit the wavefunctions to functions of the
form $\psi(y)=A\sin[p_yy+\phi_0]$, where $A$ and $\phi_0$ are constants, but inserting
phase jumps of $\pm {{\pi} \over 2}$ at the discontinuities in the wavefunctions.
Typical results are illustrated in Fig. \ref{HexRing_spectra_fit}.  One may see the
excellent fit to wavefunctions with well-defined $p_y$.  We have confirmed that
similar phase jumps and fitted wavefunctions are produced at least to the ninth energy level.

\begin{table}
\begin{tabular}{ c r r r}
%&$6(p_yL+\theta_0)=2\pi m$ & &\\
\hline

  Energy level & $p_yL$ & $\theta_0$ & m \\
  \hline 
  1 & $\pi /6$ & $\pi /2$ & 2 \\
  2 & $-\pi /6$ & $-\pi /2$ & -2 \\
  3 & $\pi /2$ & $\pi /2$ & 3 \\
  4 & $-\pi /2$ & $-\pi /2$ & -3 \\
  5 & $5\pi /6$ & $\pi /2$ & 4 \\
  6 & $-5\pi /6$ & $-\pi /2$ & -4 \\
\hline
\end{tabular}
\label{table1}
\caption{Table showing values of $p_y$, $\theta_0$, and $m$ 
in formula of form $6(p_yL+\theta_0)=2\pi m$ used in matching
numerically generated wavefunctions with forms expected from wavefunction
continuity around a ring (see text.)}
\end{table}

The results also directly support the expected sign of the phase jump, associated
with the eigenvalue of $T$.  To see this, we recognize that the wavefunctions are
linear combinations of plane waves in $y$ of equal and opposite $p_y$, such that
one obtains real eigenvectors.  The values of $p_y$ are determined by requiring
for the hexagon that $e^{(ip_yL+i\pi/2) \times 6}=1$, where $L$ is the length of
one of the hexagon sides, and $p_y>0$.  This requirement very accurately produces
the values of $p_y$ used in fitting to sine functions, such as in Fig. \ref{HexRing_spectra_fit},
as well as the energies $\varepsilon=v_F|p_y|$ (see Table I.)
The actual wavefunctions are linear combinations of $e^{\pm(ip_yy+i\pi s/2)}$, with $s$
an integer $s=0,..,5$ labeling which leg of the hexagon $y$ is on; thus we see the
phase jump must have different signs for the two signs of $p_y$.  Finally, we have
examined the states with $\varepsilon<0$, and indeed find a change in sign
for the $\pi/2$ phase jumps relative to wavefunctions with the same $p_y$ and $\varepsilon>0$.

The presence of six $\pi/2$ phase jumps (due to the six corners of the hexagon) implies that
$p_y=0$ is not an allowed momentum for the electron wavefunction in this type of ring.
Thus there is no allowed zero energy state, as would be expected for a metallic
ribbon closed into an annulus,
and its absence is imposed by the presence of phase factors that are very suggestive
of effective magnetic flux penetrating the ring, as discussed above.  This the
gap in the spectrum around $\varepsilon=0$ may be interpreted as one signature
of EBTRS.

\begin{figure}[htb]
\subfigure[Pentagon]{
\includegraphics[width=\columnwidth]{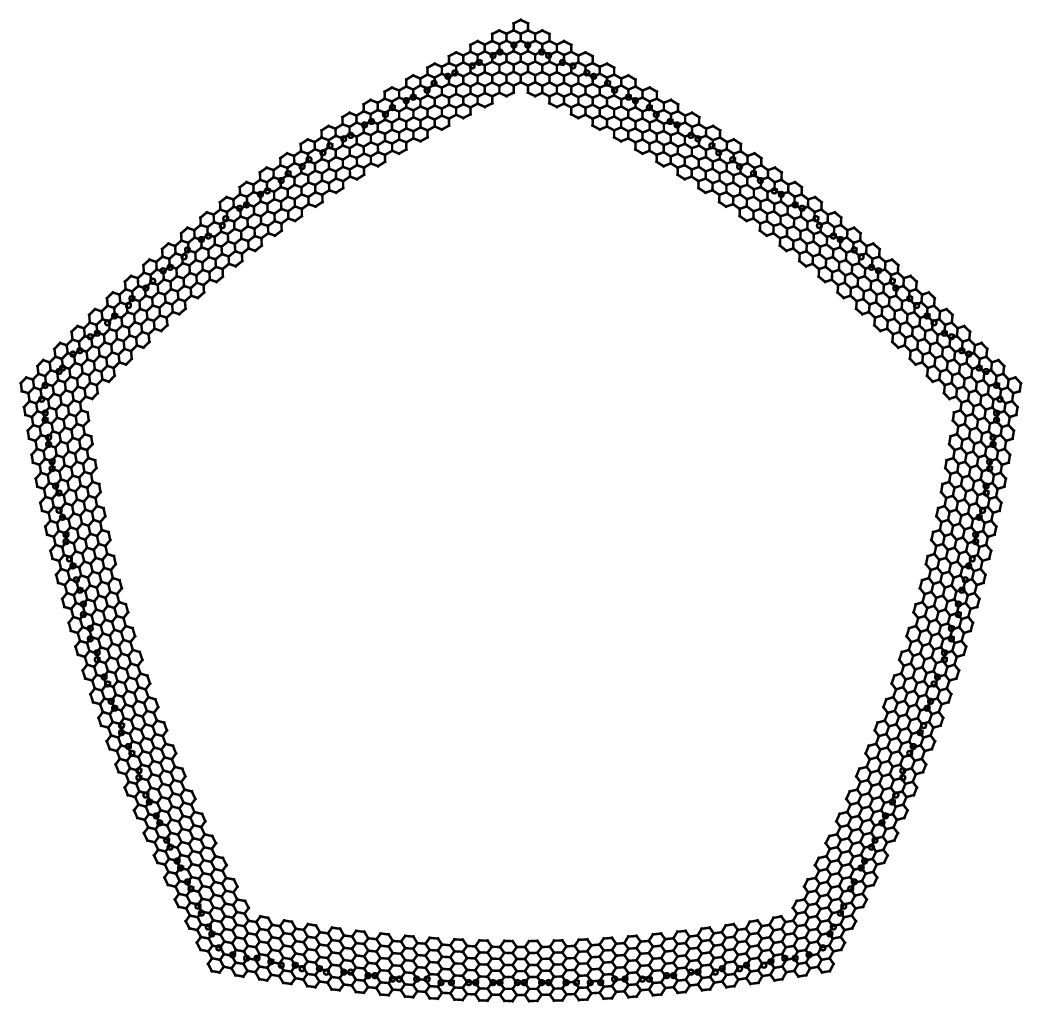}
}
\subfigure[Heptagon]{
\includegraphics[width=\columnwidth]{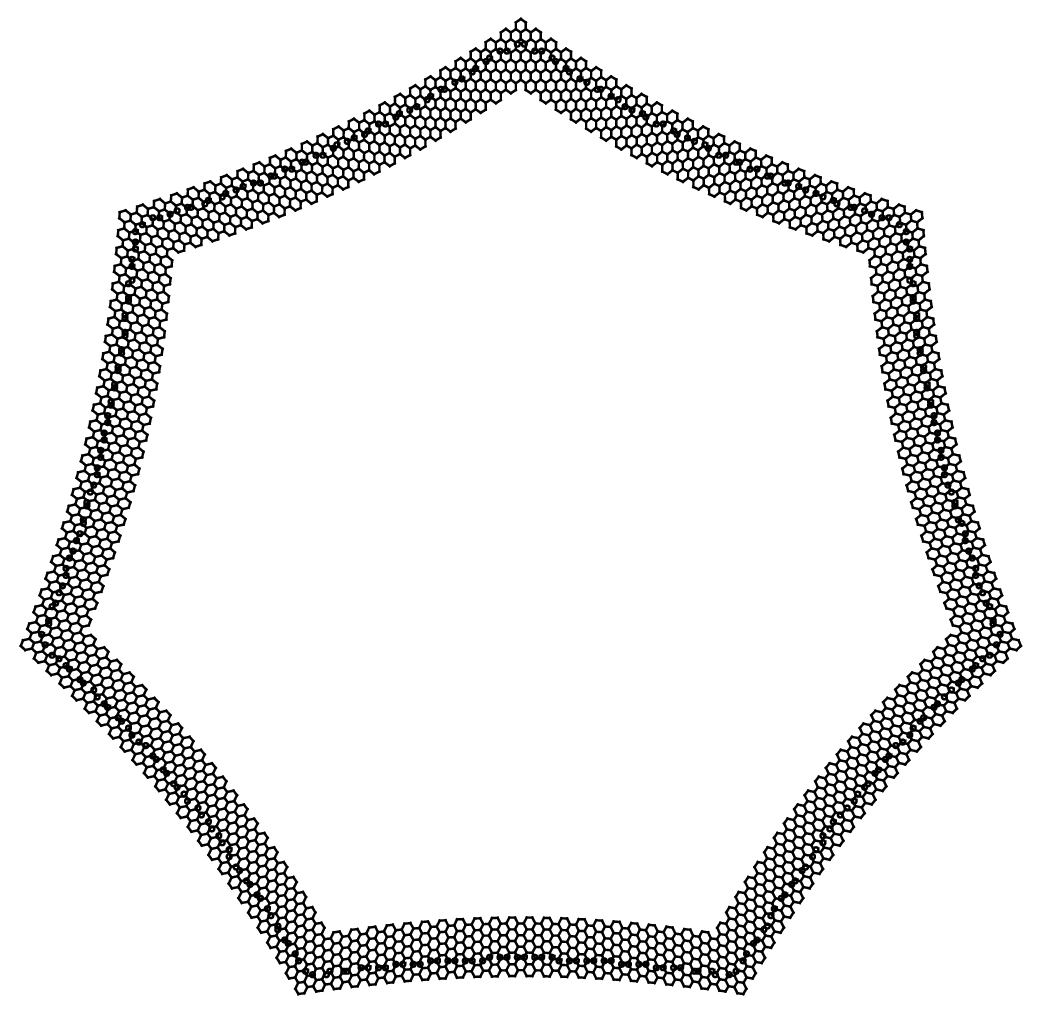}
}
\caption{Examples of (a) five-sided and (b) seven-sided graphene rings with $60^{\circ}$
corners, along with points used for analyzing wavefunctions.  Note such rings
necessarily have curvature if their bond lengths are undistorted.
}
\label{pent_and_hep}
\end{figure}

\begin{figure}[htb]
\subfigure[Pentagon]{
\includegraphics[scale=.5]{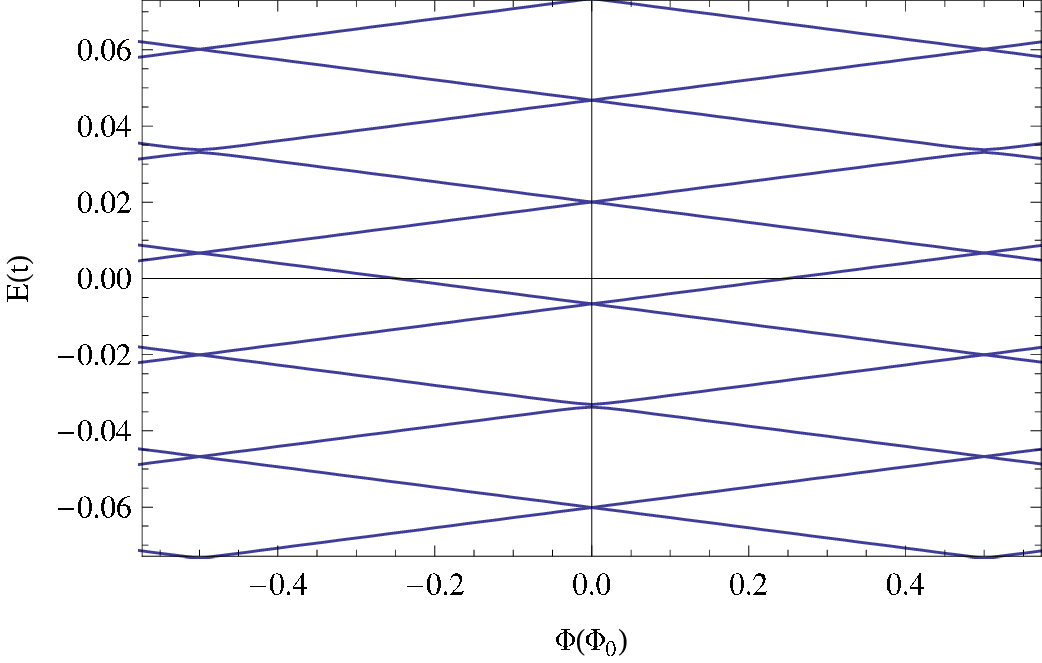}
}
\subfigure[Heptagon]{
\includegraphics[scale=.5]{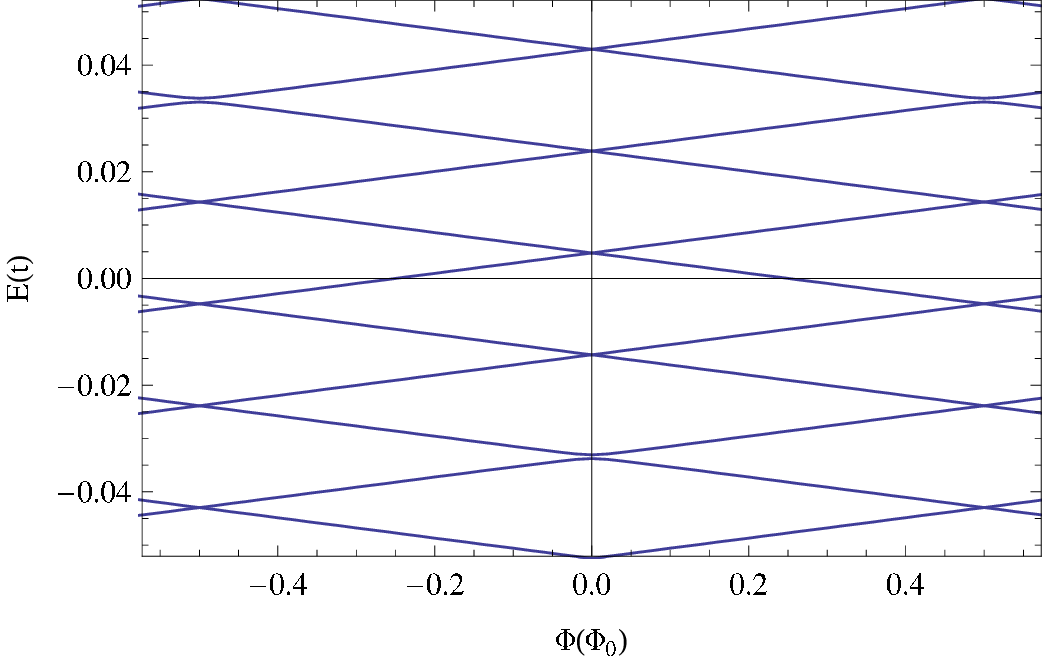}
}
\caption{Energy spectra as a function of flux for (a) pentagonal ring, and (b) heptagonal ring.
Ribbons used in constructing the polygon edges are identical to those
used for hexagonal rings with $r_a=32.5a$ and $r_b=38.5a$.}
\label{pent_and_hep_spectra}
\end{figure}

Rings with numbers of sides differing from six may also be considered, such as
those illustrated in Fig. \ref{pent_and_hep}.  Because the number of corners in these
structures is different than that of hexagons, the effective flux through
such rings will be different, leading to extra factors of $\pm i$ which
must be included in determining the wavefunctions.  These may be examined by
solving the corresponding tight-binding models, and one finds that the spectra and
wavefunctions may be fully understood in the same way as described above
for hexagonal rings.  Examples of spectra are illustrated in Fig. \ref{pent_and_hep_spectra}.
At $\Phi=0$, the energy levels precisely obey $\varepsilon=\pm v_F|p_y|$, with
$p_y$ chosen to be consistent with the total phase jump associated with the number
of corners.  Fits analogous to those shown in Fig. \ref{HexRing_spectra_fit}
may be made with equal success.

The most prominent difference between the spectra
displayed in Fig. \ref{pent_and_hep_spectra} and those of hexagonal rings is
the obvious lack of particle-hole symmetry.  In the continuum limit, particle-hole
conjugates are related by changing the relative sign of the amplitudes
on different sublattices (equivalent to multiplying the wavefunction by the Pauli matrix
$\sigma_z$ for each valley), and the resulting wavefunction has the same
value of $p_y$, but an energy of opposite sign relative to that of the original
wavefunction.  For geometries containing the junctions studied
in this work, however, there is a further effect: the phase jump for
a wavefunction with a given eigenvalue of $T$ changes sign (Eq. \ref{T_eigenvalues}.)
For even numbers of corners this has no net effect (e.g., $(\pm i)^6=-1$), but
for odd numbers this affects the wavefunction continuity condition such that
$p_y$ in general cannot take the same values for positive and negative
energies.  In terms of effective flux through the ring, this is a
reflection of the fact that the associated vector potential is a
matrix rather than a scalar quantity, allowing the system to have
different values for the effective flux for positive and negative energies.
Moreover, for a given eigenvalue of $T$, the effective flux through
a six-sided ring is effectively $\pm \Phi_0/2$, while for five- or seven-sided
rings it is $\pm\phi_0/4$.  This effective flux can be canceled by adding {\it real}
magnetic flux through the ring, bringing some states to zero energy, but the amount of
flux required depends on the number of sides [cf. Figs. \ref{HexRingAC_spectrum} and
\ref{pent_and_hep_spectra}.]

Thus, the absence of particle-hole symmetry in the spectrum may be understood
as a unique effect of the types of gauge fields which can occur in graphene.
As in the case of disclination defects \cite{Vozmediano_1993,Lammert_2000},
this effect is intimately connected to the presence of curvature in the system:
one may create pentagons and heptagons from hexagonal rings with cut-and-paste
procedures that leave the local 3-fold coordination of the bonding intact,
but in doing so curvature is necessarily introduced, and an integral number of corners
must be added or subtracted from the system.  When this number is odd, particle-hole
symmetry breaking will necessarily be present in the spectrum.  Thus, this symmetry
breaking is intimately connected to the introduction of curvature associated
with this type of topological defect.

Finally, it is interesting to consider how particle-hole symmetry may be broken from
the perspective of the tight-binding model.  It is easy to show that graphene sheets
with simple nearest neighbor hopping of uniform magnitude will support particle-hole
symmetry if all paths beginning and ending at the same point have an even number
of steps \cite{com_locator}.  Locally this is the case for all the structures
considered in this paper.  However, for rings with an odd number of sides, paths
that enclose the hole of the ring will have an {\it odd} number of steps, leading
to the breaking of particle-hole symmetry.  The fact that one must examine paths
that enclose the hole to uncover the symmetry breaking is consistent with the
idea that the effect is related to flux through the hole in the gauge-field
description.

\section{Distorted Geometries}
\label{section:perts}

In this work we have focused on ideal graphene rings in order to focus on signatures
of EBTRS in their spectra.  However, currently available fabrication techniques do not allow
the creation of such perfect rings, and an obvious question is how robust the
features discussed above might be with respect to deviations from the ideal structure.
In this section we examine this question in a few simple cases.

\begin{figure}[htb]
\subfigure[]
{
\includegraphics[scale=.5]{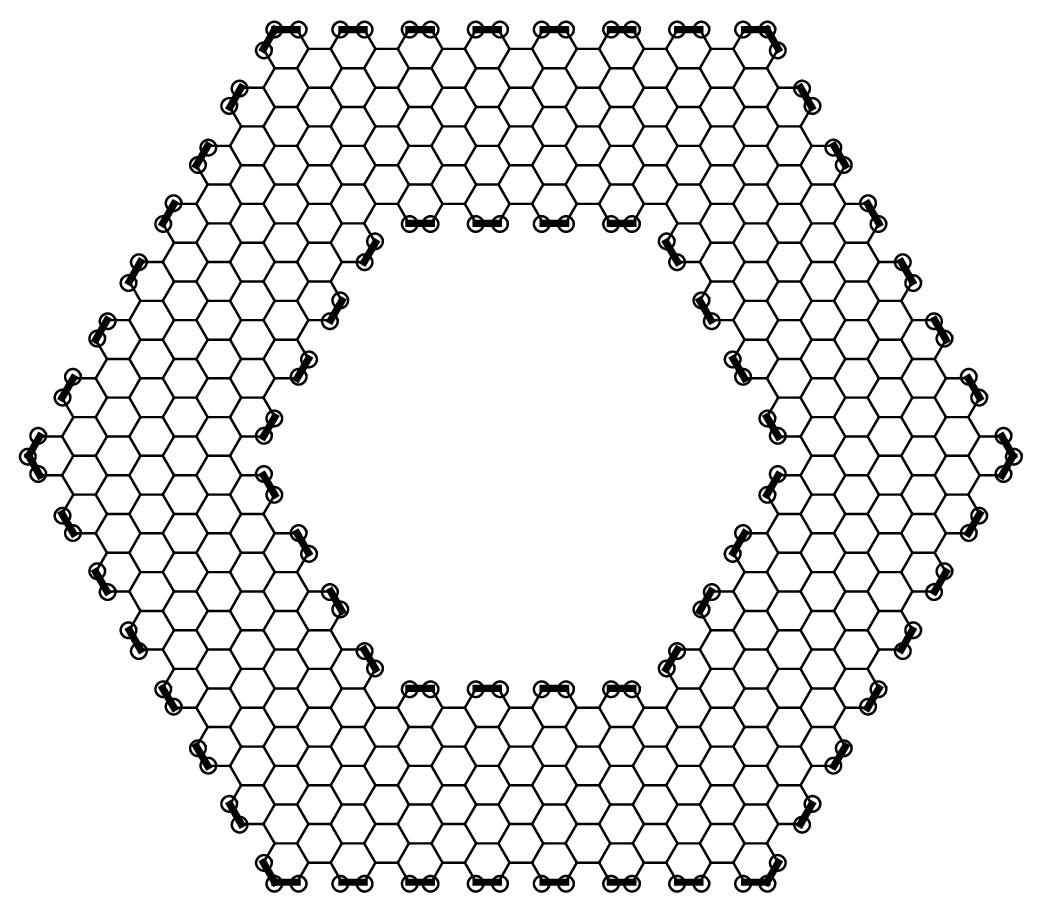}
}
\subfigure[]
{
\includegraphics[scale=.5]{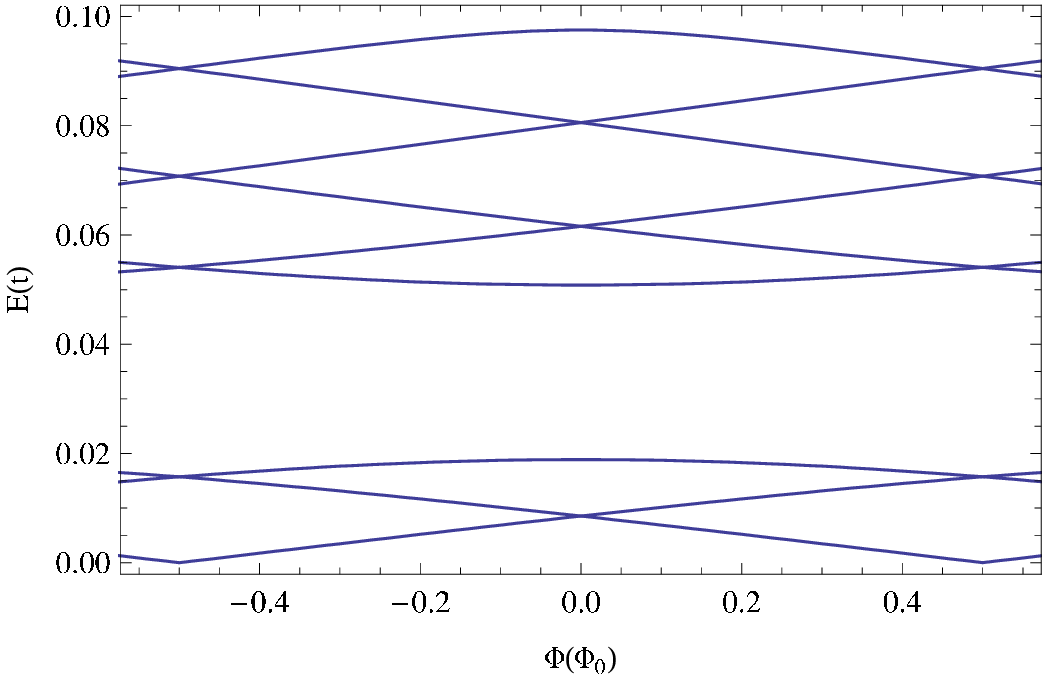}
}
\caption{Model of a hexagonal ring with passivated edges and associated energy spectrum.
(a) Bold bonds have hopping matrix elements given by 0.9$t$.
(b) Energy spectrum as function of $\Phi$.  $r_a=32.5a$, $r_b=38.5a$.}
\label{HexRing_edge}
\end{figure}

The graphene ribbon edges considered in idealized models such as considered here
have dangling bonds since atoms at the very edge only have two nearest neighbors.
In practice these bonds are often passivated with hydrogen or oxygen, and moreover
relax to slightly different bond lengths than in the bulk, so that
the gapless spectrum of ideal metallic ribbons develops a small gap \cite{Son_2006b}.
This can be modeled \cite{Son_2006b} by changing the hopping matrix element along the edge in the very last row, as illustrated in Fig. \ref{HexRing_edge}.  One may see
that the major effect of this is to open a large gap around energy $E=0.04t$, where
for the ideal ribbon this is only a small anticrossing.  Interestingly, the
gap to the first state at $\Phi=0$ is only very slightly suppressed, so that
this basic effect is unchanged by the edge modification.  Moreover, at $\Phi=\Phi_0/2$,
the levels still cross zero energy, behaving exactly as if the real flux
cancels the effective flux associated with the corners.  Thus we find that
the qualitative effects of EBTRS are robust against this edge modification.

\begin{figure}[htb]
\subfigure[]
{
\includegraphics[scale=.5]{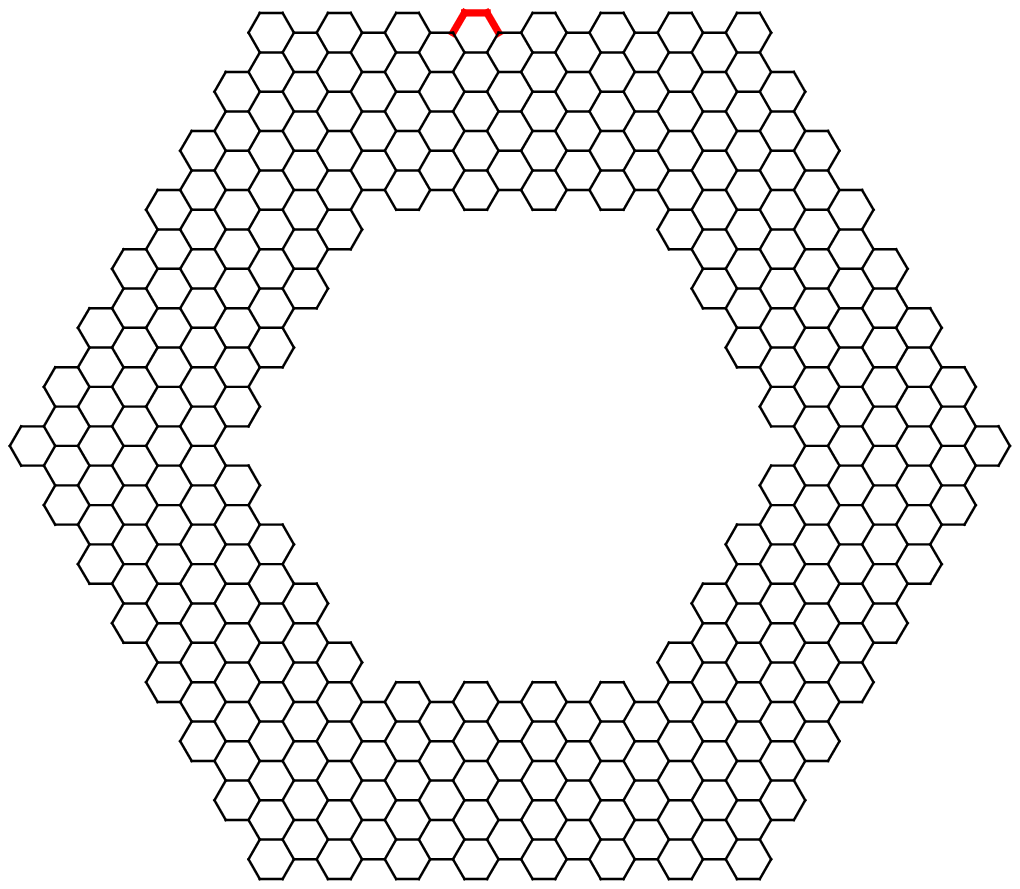}
}
\subfigure[]
{
\includegraphics[scale=.5]{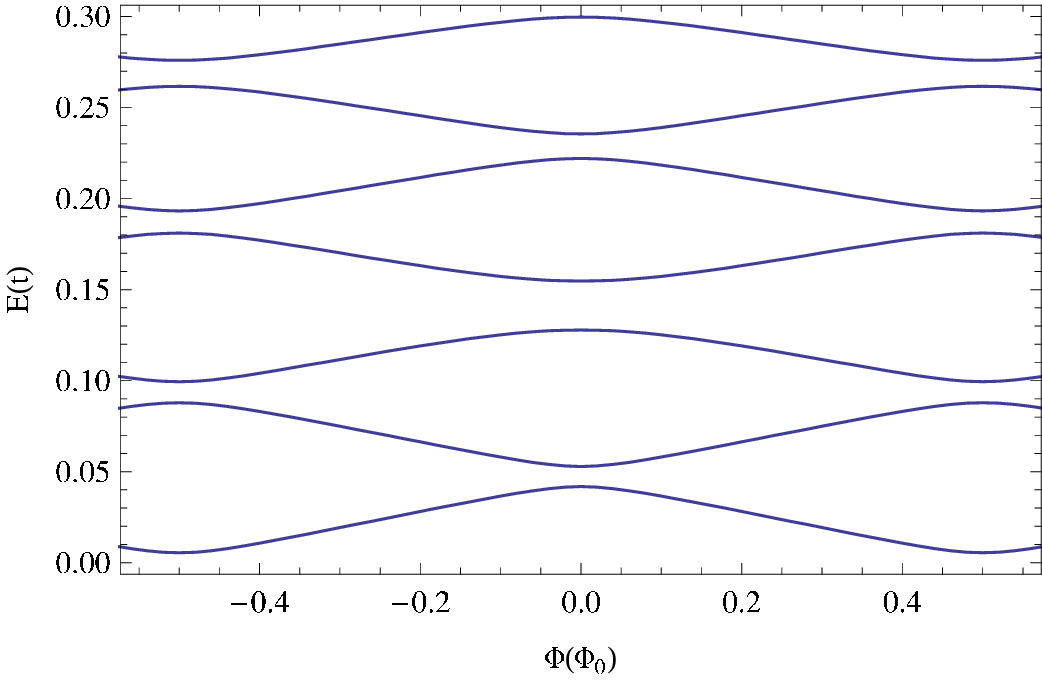}
}
\caption{(Color online.) Hexagonal ring with a single bond removed at the edge, $r_a=5.5a$, $r_b=11.5a$.
(a) Illustration of a defective bonding structure for small size system.
Bold bond and associated sites on top edge have been removed.
(b) Energy spectrum as function of $\Phi$.}
\label{HexRing_defect_1bond}
\end{figure}

\begin{figure}[htb]
\subfigure[]
{
\includegraphics[scale=.5]{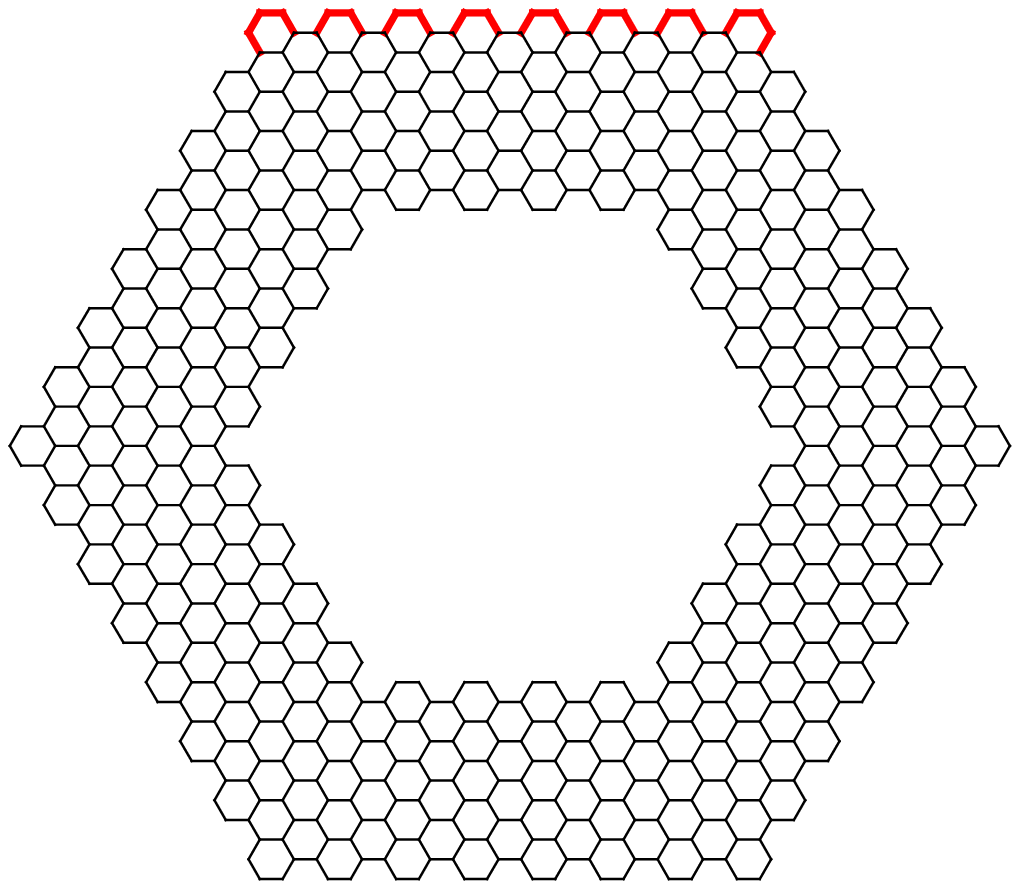}
}
\subfigure[]
{
\includegraphics[scale=.5]{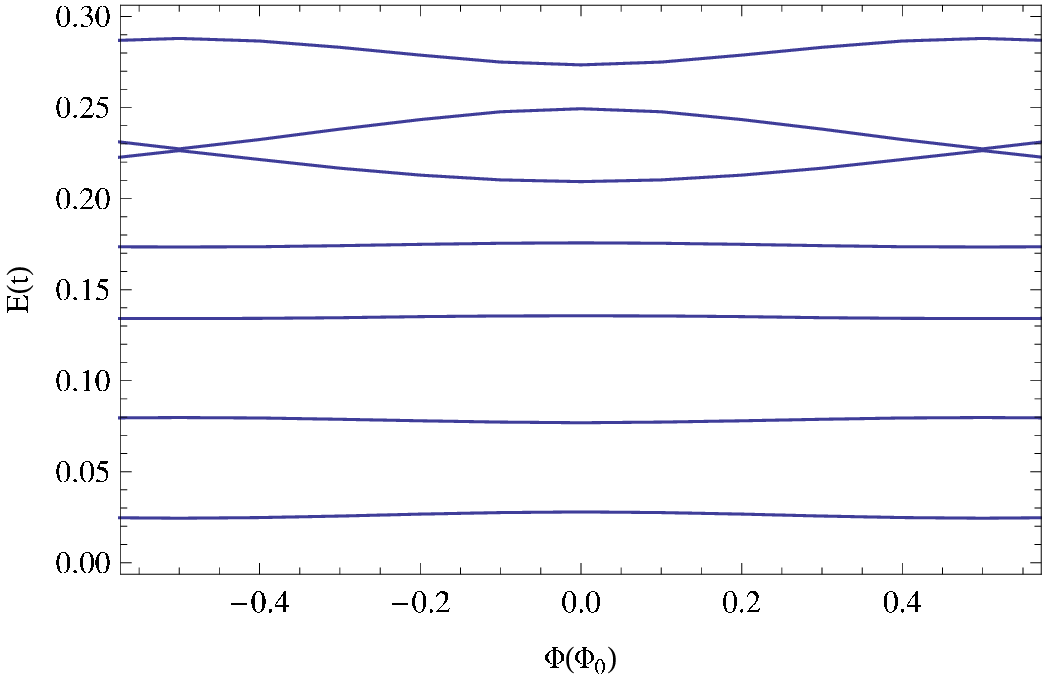}
}
\caption{(Color online.) Hexagonal ring with a row removed at the edge, $r_a=5.5a$, $r_b=11.5a$.
(a) Illustration of a defective bonding structure for small size system.
Bold bonds and associated sites of top edge have been removed.
(b) Energy spectrum as function of $\Phi$.}
\label{HexRing_defect_edge}
\end{figure}

More disruptive effects occur when the edges or corners become disordered.
Figure \ref{HexRing_defect_1bond} illustrates the effect of removing a single
pair of sites on one edge of the system.  One may see in the spectrum that
all the crossings present for the perfect system now become anticrossings.
This is easily understood: in the case of the ideal system there is perfect
six-fold symmetry, and each eigenstate carries a rotational quantum number $m$.
The perfect symmetry implies that when states with different values of $m$
approach one another as a function of $\Phi$, they cannot admix, and the
states will cross in energy.  Once the symmetry has been broken, all states can
are generically admixed, and the gap openings seen in Fig. \ref{HexRing_defect_1bond}
follow naturally.  In spite of these gaps, we see the basic structure of
the large gap around zero energy which closes as $\Phi$ approaches
$\Phi_0/2$ is maintained.

If a whole row of sites is removed, as in Fig. \ref{HexRing_defect_edge}, the spectrum
loses its essential features, and a series of states near zero energy that vary little with
$\Phi$ are apparent.  This indicates that the states have been localized; since
they do not surround the hole of the ring they are insensitive to the flux
through it.  This effect may be understood if one notes that the upper armchair
ribbon is one row narrower than the other ribbons, and hence is not conducting:
such a ribbon in isolation has a substantial gap near zero energy \cite{Brey_2006b}.
Thus we do not expect the lowest energy states to penetrate into the upper
arm of the ring, and the various effects due to any effective flux quanta
associated with the corners will be suppressed.

\begin{figure}[htb]
\subfigure[]
{
\includegraphics[scale=.5]{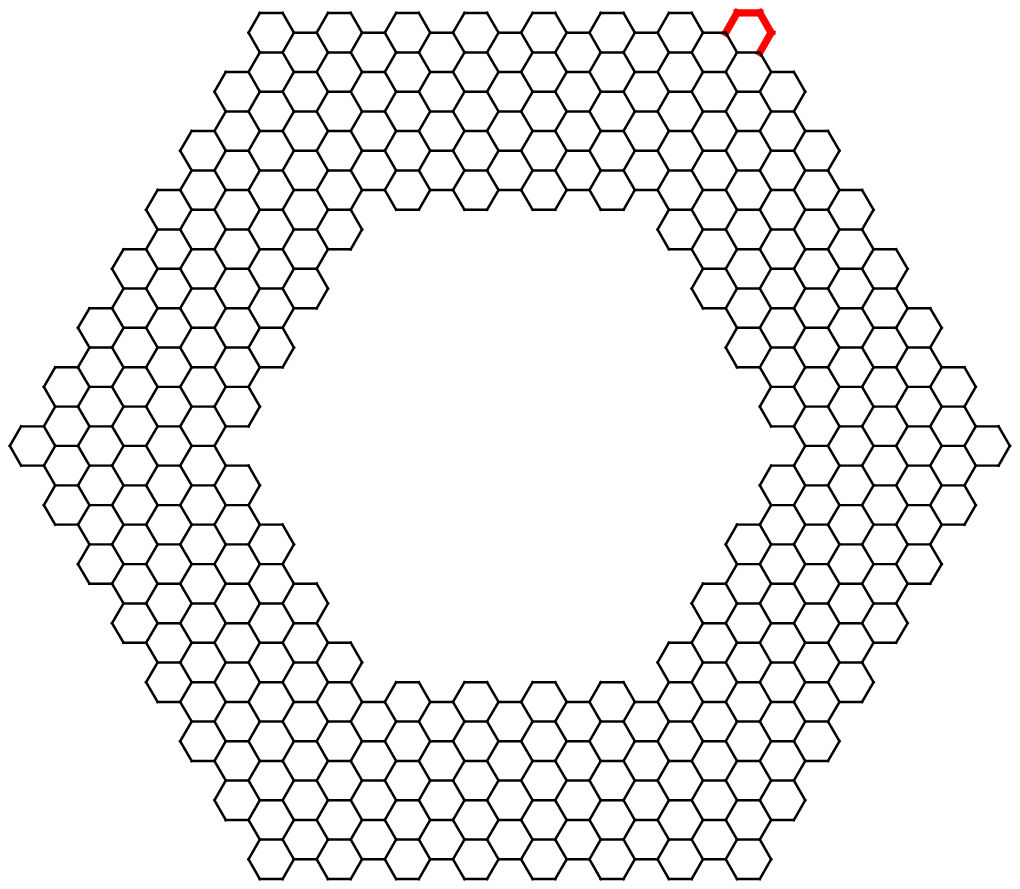}
}
\subfigure[]
{
\includegraphics[scale=.5]{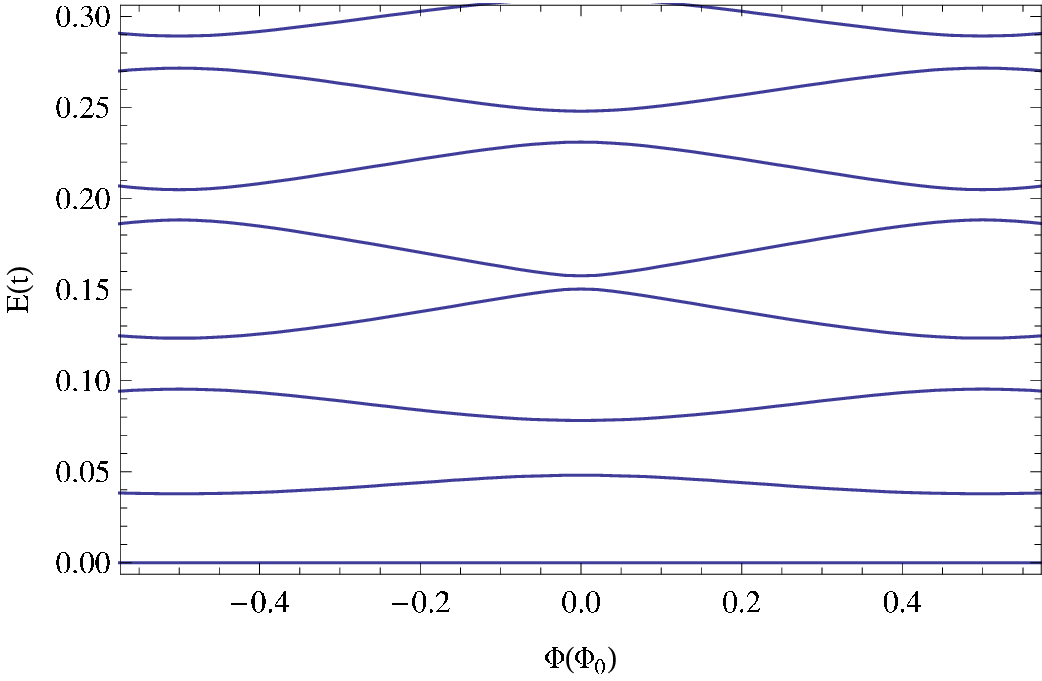}
}
\caption{(Color online.) Hexagonal ring with a site removed at one corner, $r_a=5.5a$, $r_b=11.5a$.
(a) Illustration of a defective bonding structure for small size system.
Bold bonds and associated sites have been removed.
(b) Energy spectrum as function of $\Phi$.}
\label{HexRing_defect_corner}
\end{figure}

Perhaps the most surprising effect occurs when a single site is removed from a corner,
as illustrated in Fig. \ref{HexRing_defect_corner}.  In this case one finds a non-dispersive
state precisely at zero energy.  This localized state appears to be associated with
the three point zigzag edge at the corner that results from removing this site;
analogous localized states appear in other defect structures as well
when they include such three point geometries \cite{Palacios_2008}.  
In this case the low energy structure associated with the other
corner junctions is nearly absent, suggesting the defective corner junction
has become largely reflective.  This demonstrates that the geometry of the corner
junctions plays a crucial role in the results associated with EBTRS in
graphene rings: a defective
junction will generically admix states with different $T$ eigenvalues, and moreover
can effectively block transmission at low energies so that effects
associated with the non-compactness of the geometry are difficult to
see or absent at low energies.  The sensitivity of transmission to the
precise geometry at a corner junction can also be seen in the conductance
through such junctions \cite{Iyengar_2008}.  Similar sensitivity to precise
geometries has been observed in conductance through graphene polygons and
quantum dots \cite{Iyengar_2008,Katsnelson_2008}.

We conclude this section with two remarks.  One way in which the energy spectrum
of a quantum ring is probed is through the persistent currents it supports
in a real magnetic field, $J=-c {{dE} \over {d\Phi}}$.  Such currents create
magnetization which may be detected either for a large enough
ring or for a collection of identical rings.  Since localized states
have zero or very small $|{{dE} \over {d\Phi}}|$, their presence will not be
apparent in experiments that probe the system in this way.  However they
should be visible in tunneling experiments.  Secondly, in experiments
which probe the spectrum through the magnetization of a collection of rings,
inevitably there will be variations in the precise structure of each ring,
so that few if any will have the ideal spectra of the last section.
However one may speculate that some of the features
may be reflected in the total magnetization as a function of Fermi
energy, since this involves an average over many different disorder
realizations around the ideal ring structure.  We leave this possibility
for future research.

\section{Summary}
\label{section:summary}

In this paper we have studied the states of graphene
rings formed from metallic armchair ribbons, with $60^{\circ}$
corner junctions which are perfectly transmitting at low energies.
We showed that states of the ribbons have a quantum number that
is the eigenvalue of a symmetry operator $T$, and states with the
same value of this near zero energy are connected with magnitude one through these
particular corner junctions.  However the amplitudes support
a phase jump
as one passes through the junctions, and these phase jumps
can be described in terms of a gauge field due to effective
fluxes passing just above and below a ribbon.  When such a ribbon
is closed into a ring, there is an effective flux contained such
that $p_y$, the momentum along the ribbon direction, cannot vanish
if the wavefunction is to meet appropriate boundary conditions.
As a result there are no states of a six-sided ring at zero energy in the absence
of an extermal magnetic field.
States can be driven to zero energy by application of such a
field, which effectively cancels the flux associated with
the junctions.  This behavior may be understood as a manifestation
of effective broken time reversal symmetry (EBTRS).

Five-fold and seven-fold rings enclose different total effective fluxes,
shifting the spectra such that particle-hole symmetry is broken.
This behavior is another signal of EBTRS, and moreover is intimately
connected with the curvature induced in the rings when these systems
are formed.  Analogous descriptions have been applied to graphene
sheets with disclinations \cite{Vozmediano_1993,Lammert_2000}, although
the breaking of particle-hole symmetry in the spectrum is not
so apparent in those cases as in the present one.

Finally, we examined the effects of several types of variations from the
ideal ring structure on the spectra.  A change near the edges to model
relaxation and passivation was found to only have a small quantitative
effect at the lowest energies.  Removal of a small number of
atoms from one edge opened small gaps at points where anticrossing
occurs in the ideal case, but removing a whole row rendered
one side of the ring non-conducting, effectively removing the
signature of EBTRS.  Removing a single atom from a corner introduced
a zero energy state and greatly reduced the sensitivity of the system
near zero energy to the flux through the ring. We conclude from this that
well-formed corner junctions of the type described in this
work are a necessary ingredient to see effects attributable to EBTRS.
$$
$$

\section{Acknowledgments.}

The authors acknowledge the support and hospitality of the KITP at UCSB
where part of this work was done.  This work was been
financially supported by MEC-Spain MAT2006-03741 and by the National Science
Foundation through Grant Nos. DMR-0704033 and PHY05-51164.

\end{document}